\documentclass[onecolumn]{article}
\usepackage{makeidx}
\usepackage{graphicx}
\usepackage{amsmath}

\input{tcilatex}

\begin{document}

\title{Darwinism in quantum systems?}
\author{A. Iqbal and A.H. Toor \\
Department of Electronics, Quaid-i-Azam University, \\
Islamabad 45320, Pakistan.\\
email: qubit@isb.paknet.com.pk}
\maketitle

\begin{abstract}
We investigate the role of quantum mechanical effects in the central
stability concept of evolutionary game theory i.e. an Evolutionarily Stable
Strategy (ESS). Using two and three-player symmetric quantum games we show
how the presence of quantum phenomenon of entanglement can be crucial to
decide the course of evolutionary dynamics in a population of interacting
individuals.
\end{abstract}

\section{Introduction}

Many interesting results in the recently developed quantum game theory \cite
{meyer,eisert,marinatto,jiangfeng} are about the most fundamental idea in
noncooperative game theory \cite{neumann} i.e. the Nash equilibrium (NE). A
strategy profile is a NE \cite{nash} if no player can gain by unilaterally
deviating from it. The implicit assumption behind NE is that players make
their choices simultaneously and independently. This idea also assumes that
each player participating in a game behaves rational and searches to
maximize his own payoff. In situations where evolution of complex behavior
occurs further refinements of NE concept are required \cite{weibull},
especially, when multiple NE appear in the analysis of a game. A refinement
is then needed to prefer some NE over the others. Due attention has already
been given to NE in the recent works \cite{eisert,jiangfeng} on quantum
games and this development has motivated us to study certain refinement
notions of NE in quantum games. Refinements of NE in classical game theory
are popular as well as numerous \cite{damme}. Speaking historically, the set
of refinements became so large that eventually almost any NE could be
justified in terms of someone or other's refinement \cite{weibull}.

An interesting, and fruitful as well, refinement of NE was introduced by
Maynard Smith in 1970's that became the central notion of evolutionary game
theory. In his book \textit{Evolution and the Theory of Games} \cite{smith}
he diverted attention away from elaborate definitions of rationality and
presented an evolutionary approach in classical game theory. The
evolutionary approach can be seen as a large population model of adjustment
to a NE i.e. an adjustment of population segments by evolution as opposed to
learning. Contrary to classical game theory, in evolutionary game theory the
individuals of a population, subject to evolution, are not assumed to act
consciously and rationally. Many successful applications of evolutionary
game theory appeared in mathematical biology to predict the behavior of
bacteria and insects that can hardly be said to think at all. The most
important feature of evolutionary game theory is that the assumption of
rational players, originating from classical game theory, does not remain
crucial. This important aspect appears when players' payoffs are equated to
their reproductive success.

The concept of evolutionary stability stimulated the development of
evolutionary game theory that establishes a link between game theory and the
theory of evolution. Presently the ESS theory is the central model of
evolutionary dynamics of a populations of interacting individuals. It asks,
and finds answer to it, a basic question: which states, during the course of
selection process, of a given population are stable against perturbations
induced by mutations \cite{cressman}. The ESS theory is based on Darwin's
idea of natural selection; which is shown to be describable as an algorithm
called replicator dynamic \cite{weibull}. Iterations of selections from
randomly mutating replicators is an important feature of the dynamic.
Speaking the language of game theory, the replicator dynamic says that in a
population the proportion of players which play better strategies increase
with time. When replicator dynamic is underlying process of a game the ESSs
are shown to be stable against perturbations \cite{cressman}. In other words
ESSs are, then, rest points of the replicator dynamic.

Recent developments in quantum games provide an incentive to look at the
mathematical theory of evolution, with the central idea of an ESS, in a
broader picture given by Hilbert structure of strategy space in the new
theory. This incentive is driven by a question: how game-theoretical models,
of evolutionary dynamics in a population, shape themselves in the new
settings provided to the game theory recently by quantum mechanics? Our
motivation is that quantum mechanical effects, especially entanglement, may
have decisive role in the evolutionary dynamics that have already been
successfully, and also quite rigorously, modelled using the classical game
theory. To study evolution in quantum settings we have chosen the ESS idea
mostly for simplicity and beauty of the concept. We ask questions like, how
ESSs are affected when a classical game played in a population changes
itself to one of its quantum forms? How pure and mixed ESSs are
distinguished from one another when such a change in form of a game takes
place? And most importantly, how evolutionary dynamics becomes linked to
quantum entanglement present in games that are played in quantum settings?

In earlier papers \cite{iqbal,iqbal1,iqbal2} we showed that the presence of
entanglement, in asymmetric as well as symmetric bimatrix games, can disturb
the evolutionary stability expressed by the idea of ESS. So that,
evolutionary stability of a symmetric NE can be made to appear or disappear
by controlling entanglement in symmetric and asymmetric bimatrix games. We
found a consideration of symmetric games more appropriate because the notion
of an ESS was originally investigated \cite{smith} for pair-wise symmetric
contests. In present letter we find examples of two and three-player games
where entanglement changes evolutionary stability of a symmetric NE. We
assume initial quantum states in the same, originally suggested, simpler
form presented in the scheme \cite{marinatto} that tells how to play a
quantum game. With initial states in this form, we show that two and
three-player games are distinguished in an interesting aspect i.e.
entanglement can change evolutionary stability of pure strategies in
two-player games. However, for a mixed strategy it can be done when the
number of players are increased from two to three. Before coming to quantum
settings we first describe, mathematically, the concept of an ESS in
classical evolutionary game theory.

\section{Evolutionary stability}

Maynard Smith introduced the idea of an Evolutionarily Stable Strategy (ESS)
in a seminal paper `The logic of animal conflict' \cite{smith1}. In rough
terms, an ESS is a strategy which, if played by almost all members of a
population, cannot be displaced by a small invading group playing any
alternative strategy. Suppose pairs of individuals are repeatedly drawn at
random from a large population to play a symmetric two-person game. Let the
game between the individuals be a symmetric bimatrix game represented by the
expression $G=(M,M^{T})$ where $M$ represents the payoff matrix and $T$ its
transpose. In usual notation $P(x,y)$ gives the payoff to a $x$-player
against a $y$-player for a symmetric pair-wise contest. A strategy $x$ is
said to be an ESS if for each mutant strategy $y$ there exists a positive 
\textit{invasion barrier} such that if the population share of individuals
playing the mutant strategy $y$ falls below this barrier, then $x$ earns a
higher expected payoff than $y$. Mathematically speaking, $x$ is an ESS when
for each strategy $y\neq x$ the inequality $P[x,(1-\epsilon )x+\epsilon
y]>P[y,(1-\epsilon )x+\epsilon y]$ should hold for all sufficiently small $%
\epsilon >0$; where, for example, the expression on the left-hand side is
payoff to strategy $x$ when played against the mixed strategy $(1-\epsilon
)x+\epsilon y$. This condition for an ESS can be shown \cite{smith} easily
to be equivalent to the following two requirements

\begin{eqnarray}
1.\text{ \ \ \ }P(x,x) &>&P(y,x)  \notag \\
2.\text{ If\ }P(x,x) &=&P(y,x)\ \text{then}\ P(x,y)>P(y,y)  \label{DefESS}
\end{eqnarray}

It becomes apparent that an ESS is a symmetric NE but also possesses a
stability property against mutations. Condition $1$ in the above definition
shows that $(x,x)$ is a NE for the bimatrix game $G=(M,M^{T})$ if $x$ is an
ESS. However, the reverse is not true. If $(x,x)$ is a NE, then $x$ is an
ESS only if $x$ satisfies condition $2$ in the definition (\ref{DefESS}).
The ESS condition gives a refinement on the set of symmetric Nash equilibria 
\cite{weibull}. Its essential feature is that, apart from being a symmetric
NE, it is robust against a small number of mutants appearing in a population
playing an ESS \cite{cressman}.

Now we consider quantum settings, involving two players, to investigate the
concept of evolutionary stability.

\section{Two-player case}

We use the quantization scheme suggested by Marinatto and Weber \cite
{marinatto} for the two-player quantum game of Battle of Sexes. In this
scheme the players' tactics consist of deciding the classical probabilities
of applying two unitary and Hermitian operators (the identity $I$ and the
inversion operator $C$) on an initial quantum strategy in $2\otimes 2$
dimensional Hilbert space; that can be obtained from a system of two qubits.
The tactics phase is similar to probabilistic choice between pure strategies
in the classical game theory. Interestingly, the classical form of the game
is reproduced by making the initial quantum state unentangled \cite
{marinatto}. There are some different opinions \cite{simon,marinatto1}
concerning the use of the term \textit{strategy} in Marinatto and Weber's
scheme. Earlier Eisert, Wilkins, and Lewenstein proposed a scheme \cite
{eisert} where choosing a move corresponds to a strategy. Expanding on this
work, Marinatto and Weber presented a different approach and preferred to
call \textit{tactics} the process of choosing a move when an \textit{initial
strategy}, in the form of a quantum state, is forwarded to the players. In
this paper we will refer to Marinatto and Weber's tactics as strategies or
moves and their initial strategy as initial quantum state. We then find how
entanglement affects evolutionary stability in the circumstances that a
quantum version of a game can be reduced to its classical form by removing
entanglement. Because the classical game corresponds to an unentangled
initial quantum state, a comparison between ESSs in classical and quantized
versions of the game can be made by maneuvering the initial quantum state,
in some particular form. The scheme for two-player quantum game is shown in
fig. \ref{fig. 1}.

Consider a two-player symmetric game given by the matrix

\begin{equation}
\text{Alice}^{\prime }\text{s strategy}\overset{\text{Bob's strategy}}{
\begin{array}{ccc}
& S_{1} & S_{2} \\ 
S_{1} & (\alpha ,\alpha ) & (\beta ,\gamma ) \\ 
S_{2} & (\gamma ,\beta ) & (\delta ,\delta )
\end{array}
}  \label{payoffMatrix}
\end{equation}
and played via the initial state $\left| \psi _{in}\right\rangle =a\left|
S_{1}S_{1}\right\rangle +b\left| S_{2}S_{2}\right\rangle $ where $\left|
a\right| ^{2}+\left| b\right| ^{2}=1$. A unitary and Hermitian operator $C$
used in the scheme is defined as \cite{marinatto} $C\left|
S_{1}\right\rangle =\left| S_{2}\right\rangle $, $C\left| S_{2}\right\rangle
=\left| S_{1}\right\rangle $ and $C^{\dagger }=C=C^{-1}$. Let one of the
players chooses his strategy by implementing the identity operator $I$ with
probability $p$ and the operator $C$ with probability $(1-p)$, on the
initial state $\rho _{in}$ that corresponds to $\left| \psi
_{in}\right\rangle $. Similarly, suppose the second player applies the
operators $I$ and $C$ with probabilities $q$ and $(1-q)$ respectively. The
final density matrix is written as \cite{marinatto}

\begin{equation}
\rho _{fin}=\underset{U=I,C}{\sum }\Pr (U_{A})\Pr (U_{B})[U_{A}\otimes
U_{B}\rho _{in}U_{A}^{\dagger }\otimes U_{B}^{\dagger }]
\end{equation}
where the unitary and Hermitian operator $U$ is either $I$ or $C$. $\Pr
(U_{A})$, $\Pr (U_{B})$ are the probabilities with which players $A$ and $B$
apply the operator $U$ on the initial state, respectively. It is seen that $%
\rho _{fin}$ corresponds to a convex combination of all possible quantum
operations. Payoff operators for Alice and Bob are \cite{marinatto}

\begin{equation*}
(P_{A,B})_{oper}=\alpha ,\alpha \left| S_{1}S_{1}\right\rangle +\beta
,\gamma \left| S_{1}S_{2}\right\rangle +\gamma ,\beta \left|
S_{2}S_{1}\right\rangle +\delta ,\delta \left| S_{2}S_{2}\right\rangle
\end{equation*}
The payoffs are then obtained as mean values of these operators i.e. $%
P_{A,B}=Tr\left[ (P_{A,B})_{oper}\rho _{fin}\right] $. Because the quantum
game is symmetric using the initial state $\left| \psi _{in}\right\rangle $
and the payoff matrix (\ref{payoffMatrix}), there is no need for subscripts.
We, therefore, write the payoff to a $p$ player against a $q$ player as $%
P(p,q)$. When $\overset{\star }{p}$ is a NE we find the following payoff
difference \cite{iqbal1}

\begin{gather}
P(\overset{\star }{p},\overset{\star }{p})-P(p,\overset{\star }{p})=(%
\overset{\star }{p}-p){\LARGE [}\left| a\right| ^{2}(\beta -\delta )+  \notag
\\
\left| b\right| ^{2}(\gamma -\alpha )-\overset{\star }{p}\left\{ (\beta
-\delta )+(\gamma -\alpha )\right\} {\LARGE ]}  \label{diff}
\end{gather}

\FRAME{ftbpFUw}{4.0257in}{2.8158in}{0pt}{\Qcb{The scheme to play a
two-player quantum game.}}{\Qlb{fig. 1}}{darwinism.ps}{\special{language
"Scientific Word";type "GRAPHIC";maintain-aspect-ratio TRUE;display
"ICON";valid_file "F";width 4.0257in;height 2.8158in;depth
0pt;original-width 8.3956in;original-height 6.4013in;cropleft
"0.0285";croptop "0.9052";cropright "0.9809";cropbottom "0.0339";filename
'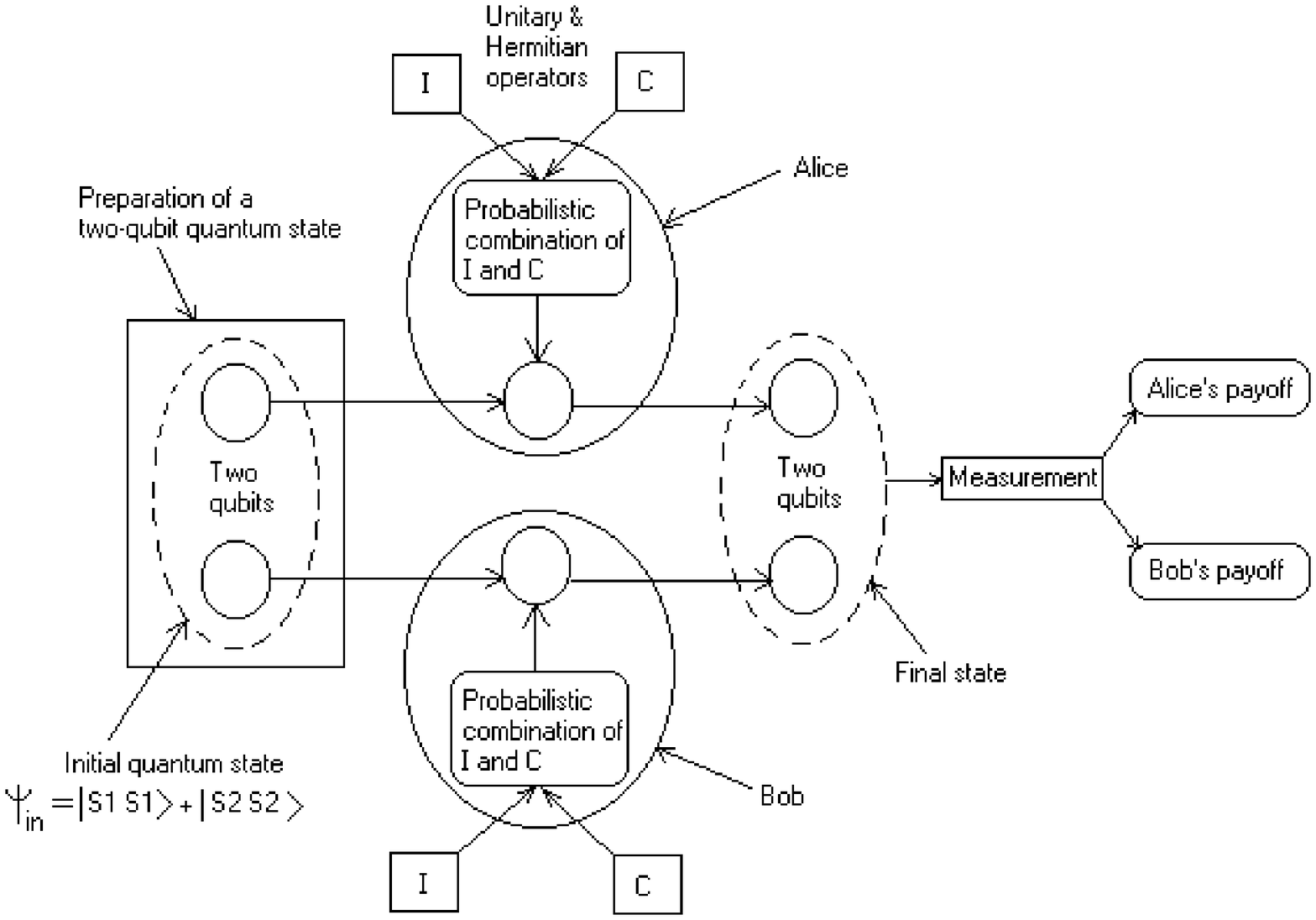';file-properties "XNPEU";}}Now the ESS conditions for the pure
strategy $p=0$ are given as

\begin{gather}
1.\text{ \ \ \ }\left| b\right| ^{2}\left\{ (\beta -\delta )-(\gamma -\alpha
)\right\} >(\beta -\delta )  \notag \\
2.\text{ If }\left| b\right| ^{2}\left\{ (\beta -\delta )-(\gamma -\alpha
)\right\} =(\beta -\delta )  \notag \\
\text{then }q^{2}\left\{ (\beta -\delta )+(\gamma -\alpha )\right\} >0
\end{gather}
where $1$ is the NE condition. Similarly the ESS conditions for the pure
strategy $p=1$ are

\begin{gather}
1.\text{ \ \ \ }\left| b\right| ^{2}\left\{ (\gamma -\alpha )-(\beta -\delta
)\right\} >(\gamma -\alpha )  \notag \\
2.\text{ If }\left| b\right| ^{2}\left\{ (\gamma -\alpha )-(\beta -\delta
)\right\} =(\gamma -\alpha )  \notag \\
\text{then }(1-q)^{2}\left\{ (\beta -\delta )+(\gamma -\alpha )\right\} >0
\end{gather}
Because these conditions for both the pure strategies $p=1$ and $p=0$ depend
on $\left| b\right| ^{2}$, therefore, there can be examples of two-player
symmetric games for which the evolutionary stability of pure strategies can
be changed while playing the game using initial state in the form $\left|
\psi _{in}\right\rangle =a\left| S_{1}S_{1}\right\rangle +b\left|
S_{2}S_{2}\right\rangle $. However, for the mixed NE, given as $\overset{%
\star }{p}=\frac{\left| a\right| ^{2}(\beta -\delta )+\left| b\right|
^{2}(\gamma -\alpha )}{(\beta -\delta )+(\gamma -\alpha )}$, the
corresponding payoff difference \ref{diff} becomes identically zero. From
the second condition of an ESS we find, for the mixed NE $\overset{\star }{p}
$, the difference

\begin{eqnarray}
&&P(\overset{\star }{p},q)-P(q,q)=\frac{1}{(\beta -\delta )+(\gamma -\alpha )%
}\times  \notag \\
&&{\LARGE [}(\beta -\delta )-q\left\{ (\beta -\delta )+(\gamma -\alpha
)\right\} -\left| b\right| ^{2}\left\{ (\beta -\delta )-(\gamma -\alpha
)\right\} {\LARGE ]}^{2}
\end{eqnarray}
Therefore, the mixed strategy $\overset{\star }{p}$ is an ESS when $\left\{
(\beta -\delta )+(\gamma -\alpha )\right\} >0$. This condition, making the
mixed NE $\overset{\star }{p}$ an ESS, is independent of $\left| b\right|
^{2}$ $\cite{remark1}$. So that, in this symmetric two-player quantum game,
evolutionary stability of the mixed NE $\overset{\star }{p}$ can not be
changed when the game is played using initial quantum states of the form $%
\left| \psi _{in}\right\rangle =a\left| S_{1}S_{1}\right\rangle +b\left|
S_{2}S_{2}\right\rangle $.

Therefore, evolutionary stability of only the pure strategies can be
affected, with the chosen form of the initial states, for the two-player
symmetric games. Examples of the games with this property are easy to find.
The class of games for which $\gamma =\alpha $ and $(\beta -\delta )<0$ the
strategies $p=0$ and $p=1$ remain NE for all $\left| b\right| ^{2}\in
\lbrack 0,1]$; but the strategy $p=1$ is not an ESS when $\left| b\right|
^{2}=0$ and the strategy $p=0$ is not an ESS when $\left| b\right| ^{2}=1$.
In an earlier letter \cite{iqbal1} we found an example of a class of games
for which a pure strategy, that is an ESS classically, does not remain ESS
for a particular value of $\left| b\right| ^{2}$,even though it remains a NE
for all possible range of $\left| b\right| ^{2}$.

To find examples of symmetric quantum games, where evolutionary stability of
the mixed strategies may also be affected by controlling the entanglement,
we now increase the number of players from two to three.

\section{Three-player case}

In extending the two-player scheme to a three-player case, we assume that
three players $A,B,$ and $C$ play their strategies by applying the identity
operator $I$ with the probabilities $p,q$ and $r$ respectively on the
initial state $\left| \psi _{in}\right\rangle $. Therefore, they apply the
operator $C$ with the probabilities $(1-p),(1-q)$ and $(1-r)$ respectively.
The final state then corresponds to the density matrix

\begin{equation}
\rho _{fin}=\underset{U=I,C}{\sum }\Pr (U_{A})\Pr (U_{B})\Pr (U_{C})\left[
U_{A}\otimes U_{B}\otimes U_{C}\rho _{in}U_{A}^{\dagger }\otimes
U_{B}^{\dagger }\otimes U_{C}^{\dagger }\right]
\end{equation}
where the $8$ basis vectors are $\left| S_{i}S_{j}S_{k}\right\rangle $, for $%
i,j,k=1,2$. Again we use initial quantum state in the form $\left| \psi
_{in}\right\rangle =a\left| S_{1}S_{1}S_{1}\right\rangle +b\left|
S_{2}S_{2}S_{2}\right\rangle $, where $\left| a\right| ^{2}+\left| b\right|
^{2}=1$. It is a quantum state in $2\otimes 2\otimes 2$ dimensional Hilbert
space that can be prepared from a system of three two-state quantum systems
or qubits. Similar to the two-player case, we define the payoff operators
for the players $A,$ $B,$ and $C$ as

\begin{eqnarray}
&&(P_{A,B,C})_{oper}=  \notag \\
&&\alpha _{1},\beta _{1},\eta _{1}\left| S_{1}S_{1}S_{1}\right\rangle
\left\langle S_{1}S_{1}S_{1}\right| +\alpha _{2},\beta _{2},\eta _{2}\left|
S_{2}S_{1}S_{1}\right\rangle \left\langle S_{2}S_{1}S_{1}\right| +  \notag \\
&&\alpha _{3},\beta _{3},\eta _{3}\left| S_{1}S_{2}S_{1}\right\rangle
\left\langle S_{1}S_{2}S_{1}\right| +\alpha _{4},\beta _{4},\eta _{4}\left|
S_{1}S_{1}S_{2}\right\rangle \left\langle S_{1}S_{1}S_{2}\right| +  \notag \\
&&\alpha _{5},\beta _{5},\eta _{5}\left| S_{1}S_{2}S_{2}\right\rangle
\left\langle S_{1}S_{2}S_{2}\right| +\alpha _{6},\beta _{6},\eta _{6}\left|
S_{2}S_{1}S_{2}\right\rangle \left\langle S_{2}S_{1}S_{2}\right| +  \notag \\
&&\alpha _{7},\beta _{7},\eta _{7}\left| S_{2}S_{2}S_{1}\right\rangle
\left\langle S_{2}S_{2}S_{1}\right| +\alpha _{8},\beta _{8},\eta _{8}\left|
S_{2}S_{2}S_{2}\right\rangle \left\langle S_{2}S_{2}S_{2}\right|
\end{eqnarray}
where $\alpha _{l},\beta _{l},\eta _{l}$ for $1\leq l\leq 8$ are $24$
constants of the matrix of this three-player game. Payoffs to the players $%
A,B,$ and $C$ are then obtained as mean values of these operators

\begin{equation}
P_{A,B,C}(p,q,r)=Trace\left[ (P_{A,B,C})_{oper}\rho _{fin}\right]
\end{equation}
Here, similar to two player case, the classical payoffs can be obtained by
making the initial quantum state unentangled and fixing $\left| b\right|
^{2}=0$. To get a symmetric game we define $P_{A}(x,y,z)$ as the payoff to
player $A$ when players $A$, $B$, and $C$ play the strategies $x$,$y$ and $z$
respectively. Following relations make payoffs to the players a quantity
that is identity independent but depends only on their strategies

\begin{gather}
P_{A}(x,y,z)=P_{A}(x,z,y)=P_{B}(y,x,z)  \notag \\
=P_{B}(z,x,y)=P_{C}(y,z,x)=P_{C}(z,y,x)
\end{gather}
For these relations to hold we need following replacements for $\beta _{i}$
and $\eta _{i}$

\begin{eqnarray}
\beta _{1} &\rightarrow &\alpha _{1}\qquad \beta _{2}\rightarrow \alpha
_{3}\qquad \beta _{3}\rightarrow \alpha _{2}\qquad \beta _{4}\rightarrow
\alpha _{3}  \notag \\
\beta _{5} &\rightarrow &\alpha _{6}\qquad \beta _{6}\rightarrow \alpha
_{5}\qquad \beta _{7}\rightarrow \alpha _{6}\qquad \beta _{8}\rightarrow
\alpha _{8}  \notag \\
\eta _{1} &\rightarrow &\alpha _{1}\qquad \eta _{2}\rightarrow \alpha
_{3}\qquad \eta _{3}\rightarrow \alpha _{3}\qquad \eta _{4}\rightarrow
\alpha _{2}  \notag \\
\eta _{5} &\rightarrow &\alpha _{6}\qquad \eta _{6}\rightarrow \alpha
_{6}\qquad \eta _{7}\rightarrow \alpha _{5}\qquad \eta _{8}\rightarrow
\alpha _{8}
\end{eqnarray}
Also, it is now necessary that we should have $\alpha _{6}=\alpha _{7}$ and $%
\alpha _{3}=\alpha _{4}.$ A symmetric game between three players, therefore,
can be defined by only six constants. We take these to be $\alpha
_{1},\alpha _{2},\alpha _{3},\alpha _{5},\alpha _{6}$ and, $\alpha _{8}$.
The payoff to a player now becomes only a strategy dependent quantity and
becomes identity independent. No subscripts are therefore, needed. Payoff to
a $p$ player, when other two players play $q$ and $r$, can now be written as 
$P(p,q,r)$. A symmetric NE $\overset{\star }{p}$ can be found from the Nash
condition $P(\overset{\star }{p},\overset{\star }{p},\overset{\star }{p}%
)-P(p,\overset{\star }{p},\overset{\star }{p})\geq 0$ i.e.

\begin{gather}
P(\overset{\star }{p},\overset{\star }{p},\overset{\star }{p})-P(p,\overset{%
\star }{p},\overset{\star }{p})=(\overset{\star }{p}-p)\text{{\LARGE [}}%
\overset{\star }{p}^{2}(1-2\left| b\right| ^{2})(\sigma +\omega -2\eta )+ 
\notag \\
2\overset{\star }{p}\left\{ \left| b\right| ^{2}(\sigma +\omega -2\eta
)-\omega +\eta \right\} +\left\{ \omega -\left| b\right| ^{2}(\sigma +\omega
)\right\} \text{{\LARGE ]}}\geq 0
\end{gather}
where ($\alpha _{1}-\alpha _{2})=\sigma $, ($\alpha _{3}-\alpha _{6})=\eta $%
, and ($\alpha _{5}-\alpha _{8})=\omega $. Three possible NE are found as

\begin{equation}
\QATOPD\{ \} {\overset{\star }{p}_{1}=\frac{\left\{ (\omega -\eta )-\left|
b\right| ^{2}(\sigma +\omega -2\eta )\right\} \pm \sqrt{\left\{ (\sigma
+\omega )^{2}-(2\eta )^{2}\right\} \left| b\right| ^{2}(1-\left| b\right|
^{2})+(\eta ^{2}-\sigma \omega )}}{(1-2\left| b\right| ^{2})(\sigma +\omega
-2\eta )}}{
\begin{array}{c}
\overset{\star }{p}_{2}=0 \\ 
\overset{\star }{p}_{3}=1
\end{array}
}
\end{equation}
Clearly the mixed NE $\overset{\star }{p_{1}}$ makes the difference $P(%
\overset{\star }{p},\overset{\star }{p},\overset{\star }{p})-P(p,\overset{%
\star }{p},\overset{\star }{p})$ identically zero and two values for $%
\overset{\star }{p}_{1}$ can be found for a given $\left| b\right| ^{2}$. $%
\overset{\star }{p}_{2}$, $\overset{\star }{p}_{3}$are pure strategy NE. We
notice that $\overset{\star }{p}_{1}$ comes out as a NE without imposing
further restrictions on the matrix of the symmetric three-player game.
However, the pure strategies $\overset{\star }{p}_{2}$and $\overset{\star }{p%
}_{3}$can be NE when further restriction are imposed on the matrix of the
game. For example, $\overset{\star }{p}_{3}$can be a NE provided $\sigma
\geq (\omega +\sigma )\left| b\right| ^{2}$ for all $\left| b\right| ^{2}\in
\lbrack 0,1]$. Similarly $\overset{\star }{p}_{2}$can be NE when $\omega
\leq (\omega +\sigma )\left| b\right| ^{2}$.

Now the question we ask: how the evolutionary stability of these three NE
can be affected while playing the game via the initial quantum states given
in the form $\left| \psi _{in}\right\rangle =a\left|
S_{1}S_{1}S_{1}\right\rangle +b\left| S_{2}S_{2}S_{2}\right\rangle $?. For
the two-player asymmetric game of Battle of Sexes we showed that out of the
three NE only two can be evolutionarily stable \cite{iqbal}. In classical
evolutionary game theory the concept of an ESS is well known to be extended
to multi-player case. When mutants are allowed to play only one strategy the
definition of an ESS for three-player case is written as \cite{mark}

\begin{eqnarray}
1.\text{ \ \ \ }P(p,p,p) &>&P(q,p,p)  \notag \\
2.\text{ If }P(p,p,p) &=&P(q,p,p)\text{ then }P(p,q,p)>P(q,q,p)
\end{eqnarray}
Here $p$ is a NE if it satisfies the condition $1$ against all $q\neq p$.
For our case the ESS conditions for the pure strategies $\overset{\star }{p}%
_{2}$and $\overset{\star }{p}_{3}$ can be written as follows. For example, $%
\overset{\star }{p}_{2}=0$ is an ESS when

\begin{eqnarray}
1.\text{ \ \ \ }\sigma \left| b\right| ^{2} &>&\omega \left| a\right| ^{2} 
\notag \\
2.\text{ If }\sigma \left| b\right| ^{2} &=&\omega \left| a\right| ^{2}\text{
then }-\eta q^{2}(\left| a\right| ^{2}-\left| b\right| ^{2})>0
\end{eqnarray}
where $1$ is NE condition for the strategy $\overset{\star }{p}_{2}=0$.
Similarly $\overset{\star }{p}_{3}=1$ is an ESS \cite{remark2} when

\begin{eqnarray}
1.\text{ \ \ \ }\sigma \left| a\right| ^{2} &>&\omega \left| b\right| ^{2} 
\notag \\
2.\text{ If }\sigma \left| a\right| ^{2} &=&\omega \left| b\right| ^{2}\text{
then }\eta (1-q)^{2}(\left| a\right| ^{2}-\left| b\right| ^{2})>0
\end{eqnarray}
and both the pure strategies $\overset{\star }{p}_{2}$and $\overset{\star }{p%
}_{3}$ are ESSs when $\left| a\right| ^{2}=\left| b\right| ^{2}$. Examples
of three-player symmetric games are easy to find for which a pure strategy
is a NE for the whole range $\left| b\right| ^{2}\in \lbrack 0,1]$, but does
not remain an ESS for some particular value of $\left| b\right| ^{2}$. An
example of a class of such games is for which $\sigma =0,\omega <0$ and $%
\eta \leq 0$. In this class the strategy $\overset{\star }{p}_{2}=0$ is a NE
for all $\left| b\right| ^{2}\in \lbrack 0,1]$ but not an ESS when $\left|
b\right| ^{2}=1$.

However, the mixed strategy NE $\overset{\star }{p}_{1}$ forms the most
interesting case. It makes the payoff difference $P(\overset{\star }{p_{1}},%
\overset{\star }{p_{1}},\overset{\star }{p_{1}})-P(p,\overset{\star }{p_{1}},%
\overset{\star }{p_{1}})$ identically zero for every strategy $p$. Now $%
\overset{\star }{p_{1}}$ is an ESS when $\left\{ P(\overset{\star }{p_{1}},q,%
\overset{\star }{p_{1}})-P(q,q,\overset{\star }{p_{1}})\right\} >0$ but

\begin{eqnarray}
&&P(\overset{\star }{p_{1}},q,\overset{\star }{p_{1}})-P(q,q,\overset{\star 
}{p_{1}})  \notag \\
&=&\pm (\overset{\star }{p_{1}}-q)^{2}\sqrt{\left\{ (\sigma +\omega
)^{2}-(2\eta )^{2}\right\} \left| b\right| ^{2}(1-\left| b\right|
^{2})+(\eta ^{2}-\sigma \omega )}  \label{sqr}
\end{eqnarray}
Therefore, out of the two possible roots $(\overset{\star }{p_{1}})_{1}$ and 
$(\overset{\star }{p_{1}})_{2}$, that make the difference $P(\overset{\star 
}{p_{1}},q,\overset{\star }{p_{1}})-P(q,q,\overset{\star }{p_{1}})$ greater
than and less than zero respectively, of the quadratic equation

\begin{gather}
\overset{\star }{p_{1}}^{2}(1-2\left| b\right| ^{2})(\sigma +\omega -2\eta )+
\notag \\
2\overset{\star }{p_{1}}\left\{ \left| b\right| ^{2}(\sigma +\omega -2\eta
)-\omega +\eta \right\} +\left\{ \omega -\left| b\right| ^{2}(\sigma +\omega
)\right\} =0
\end{gather}
only $(\overset{\star }{p_{1}})_{1}$ can exist as an ESS. When the square
root term in the equation (\ref{sqr}) becomes zero we have only one mixed
NE, that is not an ESS. Therefore, out of four possible NE in this
three-player game only three can be ESSs. An interesting class of
three-player games is one with $\eta ^{2}=\sigma \omega $. For these games
the mixed NE are $\overset{\star }{p_{1}}=\frac{\left\{ (w-\eta )-\left|
b\right| ^{2}(\sigma +\omega -2\eta )\right\} \pm \left| a\right| \left|
b\right| \left| \sigma -\omega \right| }{(1-2\left| b\right| ^{2})(\sigma
+\omega -2\eta )}$ and, when played classically, we can get only one mixed
NE that is not an ESS. However for all $\left| b\right| ^{2}$, different
from zero, we generally obtain two NE out of which one can be an ESS.

Similar to the two-player case, the NE in a three-player symmetric game
important from the point of view of evolutionary stability are those that
survive a change between two initial states; one being unentangled
corresponding to the classical game. Suppose $\overset{\star }{p_{1}}$
remains a NE for $\left| b\right| ^{2}=0$ and some other non-zero $\left|
b\right| ^{2}$. It is possible when $(\sigma -\omega )(2\overset{\star }{%
p_{1}}-1)=0$. One possibility is the strategy $\overset{\star }{p}=\frac{1}{2%
}$ remaining a NE for all $\left| b\right| ^{2}\in \lbrack 0,1]$. It reduces
the defining quadratic equation for $\overset{\star }{p_{1}}$ to $\sigma
+\omega +2\eta =0$ and makes the difference $P(\overset{\star }{p_{1}},q,%
\overset{\star }{p_{1}})-P(q,q,\overset{\star }{p_{1}})$ independent of $%
\left| b\right| ^{2}$. Therefore the strategy $\overset{\star }{p}=\frac{1}{2%
},$ even when remaining a NE for all $\left| b\right| ^{2}\in \lbrack 0,1],$
can not be an ESS in only one version of the symmetric three-player game.
For the second possibility $\sigma =\omega $ the defining equation for $%
\overset{\star }{p_{1}}$is reduced to

\begin{equation}
(1-2\left| b\right| ^{2})\left\{ \overset{\star }{p_{1}}-\frac{(\eta -\sigma
)-\sqrt{\eta ^{2}-\sigma ^{2}}}{2(\eta -\sigma )}\right\} \left\{ \overset{%
\star }{p_{1}}-\frac{(\eta -\sigma )+\sqrt{\eta ^{2}-\sigma ^{2}}}{2(\eta
-\sigma )}\right\} =0
\end{equation}
for which

\begin{equation}
P(\overset{\star }{p_{1}},q,\overset{\star }{p_{1}})-P(q,q,\overset{\star }{%
p_{1}})=\pm 2(\overset{\star }{p_{1}}-q)^{2}\left| \left| b\right| ^{2}-%
\frac{1}{2}\right| \sqrt{\eta ^{2}-\sigma ^{2}}
\end{equation}
Here the difference $P(\overset{\star }{p_{1}},q,\overset{\star }{p_{1}}%
)-P(q,q,\overset{\star }{p_{1}})$ still depends on $\left| b\right| ^{2}$
and becomes zero for $\left| b\right| ^{2}=\frac{1}{2}$. Therefore for the
class of games for which $\sigma =\omega $ and $\eta >\sigma $, one of the
mixed strategies $(\overset{\star }{p_{1}})_{1},(\overset{\star }{p_{1}})_{2}
$ remains a NE for all $\left| b\right| ^{2}\in \lbrack 0,1]$ but not an ESS
when $\left| b\right| ^{2}=\frac{1}{2}$. In this class of three-player
quantum games the evolutionary stability of a mixed strategy can, therefore,
be changed while the game is played using initial quantum states in the form 
$\left| \psi _{in}\right\rangle =a\left| S_{1}S_{1}S_{1}\right\rangle
+b\left| S_{2}S_{2}S_{2}\right\rangle $.

\section{Discussion}

The fact that classical games are played in natural macroscopic world is
well known for a long time. Evolutionary game theory is a subject, growing
out of such studies and, dealing mostly with games played in the animal
world. Recent work in biology \cite{turner} suggests nature playing
classical games at micro-level. Bacterial infections by viruses have been
presented as classical game-like situations where nature prefers the
dominant strategies. The concept of evolutionary stability, without assuming
rational and conscious individuals, gives game-theoretical models of stable
states for a population of interacting individuals. Darwinian idea of
natural selection provides physical ground to these models of rationality
studied in evolutionary game theory. Why there is need to study evolutionary
stability in quantum games? We find it interesting that some entirely
quantum aspect like entanglement can have a deciding role about which stable
states of the population should survive and others should not. It means that
the presence of quantum interactions, in a population undergoing evolution,
can alter its stable states resulting from evolutionary dynamics. When
entanglement decides the evolutionary outcomes, the role for quantum
mechanics clearly increases from just keeping atoms together that constitute
molecules. This new role may now also include to define and maintain
complexity emerging from quantum interactions among a collection of
molecules. It becomes even more interesting with reference to the example of
equilibrium in a mixture of chemicals presented above. When quantum nature
of molecular interactions decides the equilibria that the mixture of
Schuster et al. \cite{schuster} should be able to attain, there is a clear
possibility for the quantum mechanical role in the models of
self-organization in matter and the evolution of macromolecules before the
advent of life.

We have two suggestions where this finding, about quantum effects deciding
evolutionary outcomes in a population of interacting entities, can have a
relevance

\subsection{Genetic code evolution}

The genetic code is the relationship between the sequence of the bases in
the DNA and the sequence of amino acids in proteins. Recent work \cite
{knight} about evolvability of the genetic code suggests that the code, like
all other features of organisms, was shaped by natural selection. The
question about the process and evolutionary mechanism by which the genetic
code was optimized is still unanswered. Two major suggested possibilities
are \textit{(a).} A large number of codes existed out of which the adaptive
one was selected. \textit{(b).} Adaptive and error-minimizing constraints
gave rise to an adaptive code via code expansion and simplification. The
second possibility of code expansion from earlier simpler forms is now
thought to be supported by much empirical and genetic evidence \cite{knight1}
and results suggest that the present genetic code was strongly influenced by
natural selection for error minimization. Recently Patel \cite{patel}
suggested quantum dynamics played a role in the DNA replication and the
optimization criteria involved in genetic information processing. He
considers the criteria involved as a task similar to an unsorted assembly
operation where the Grover's database search algorithm \cite{grover}
fruitfully applies; given the different optimal solutions for classical and
quantum dynamics. The assumption underlying this approach, as we understood
it, is that an adaptive code was selected out of a large numbers that
existed earlier. Recent suggestions about natural selection being the
process for error minimization in the mechanism of adaptive code evolution
suggests, instead, an evolutionary approach for this optimization problem.
We suggest that, in the evolution and expansion of the code from its earlier
simpler forms, quantum dynamics too has played important role. The
mechanism, however, leading now to this optimization will be completely
different. Our result that stable outcomes, of an evolutionary process based
on natural selection, depend on presence or absence of quantum entanglement
clearly implies the possibility that other quantum interactions may also
have deciding role in obtaining some optimal outcome of evolution for a
system of molecules constituting a population. The mathematically rigorous
representation of stability in an evolutionary dynamics, based on Darwinian
selection, is the concept of an ESS. We believe that the code optimization
is a problem having close similarities with the problem of evolutionary
stability. By showing that entanglement can bring or take away evolutionary
stability from a symmetric NE we have indicated that the code optimization
was probably achieved by forces and effects that were quantum mechanical in
nature.

\subsection{Quantum evolutionary algorithms}

A polynomial time algorithm that can solve an NP problem is not known yet. A
viable alternative approach, shown to find acceptable solutions within a
reasonable time period, is the evolutionary search \cite{back}. Iteration of
selection based on competition, random variation usually called mutation,
and exploration of the fitness landscape of possible solutions are the basic
ingredients of many distinct paradigms of evolutionary computing \cite{back1}%
. On the other hand superposition of all possible solution states, unitary
operator exploiting interference to enhance the amplitude of the desired
states, and final measurement extracting the solution are the components of
quantum computing. These two approaches in computing are believed to
represent different philosophies \cite{greenwood}. Finding ESSs can be
easily formulated as an evolutionary algorithm with mutations occurring
within only a small proportion of the total population. In fact ESSs also
constitute an important technique in evolutionary computation. Our proposal
that entanglement has a role in the theory of ESSs suggests that the two
philosophies, considered different, may have some common grounds that
possibly unites them. It also hints the possibility of other evolutionary
algorithms that utilize or even exploit quantum effects. In an evolutionary
algorithm, exploiting quantum effects, we may have, for example, fitness
functions depending on the amount of entanglement present. The interesting
question then is: how the population will evolve towards a solution or an
equilibrium in relation to the entanglement?

\section{Conclusion}

In this paper we have shown that the idea of evolutionary stability, the
central stability concept of evolutionary game theory, can be related to
quantum entanglement. We investigated ESSs in three-player quantum games and
compared them to two-player games played by a proposed scheme where two,
Hermitian and unitary, operators are applied on an initial quantum state
with classical probabilities. We used the initial quantum state in the same
form proposed by Marinatto and Weber \cite{marinatto}. In two-player
symmetric games we found that evolutionary stability of a mixed strategy can
not be changed by a unitary maneuver of the initial quantum state. However,
for a class of three-player symmetric games it becomes possible to do so. It
shows that the presence of quantum mechanical effects may have a deciding
role on the outcomes of evolutionary dynamics in a population of interacting
entities. We suggested that a relevance of these ideas may be found, for
example, in the studies of the evolution of genetic code at the dawn of
life. Another suggestion is designing evolutionary algorithms where
interactions between individual of a population are governed by quantum
effects. The nature of these quantum effects, influencing the course of
evolution, will also determine the evolutionary outcome. Quantum mechanics
playing a role in the theory of ESSs implies that Darwin's idea of natural
selection has a relevance even for quantum systems.

\section{Acknowledgment}

This work is supported by Pakistan Institute of Lasers and Optics, Islamabad.


\begin{thebibliography}{99}
\bibitem{remark1}  An alternative possibility is that we adjust $\left|
b\right| ^{2}$=$\frac{(\beta -\sigma )-q\left\{ (\beta -\sigma )+(\gamma
-\alpha )\right\} }{\left\{ (\beta -\sigma )-(\gamma -\alpha )\right\} }$
and it makes $P(\overset{\star }{p},q)-P(q,q)$ zero and the mixed strategy $%
\overset{\star }{p}$ does not remain an ESS. However such `mutant dependent'
adjustment of $\left| b\right| ^{2}$ is not reasonable because the mutant
strategy $q$ can be anything in the range $[0,1]$.
\bibitem{remark2}  These conditions required to make the strategy $\overset{%
\star }{p}_{3}=1$ an ESS can also be written as: (1): $\sigma >(\omega
+\sigma )\left| b\right| ^{2}$ (2): If $\sigma =\left| b\right| ^{2}(\omega
+\sigma )$ then $\frac{\gamma (\omega -\sigma )}{(\omega +\sigma )}>0$. For
the strategy $\overset{\star }{p}_{2}=0$ the ESS conditions reduce to: $(1).$
$\omega <(\omega +\sigma )\left| b\right| ^{2}$ $(2).$ If $\omega =\left|
b\right| ^{2}(\omega +\sigma )$ then $\frac{\gamma (\omega -\sigma )}{%
(\omega +\sigma )}>0$.
\end{thebibliography}
\end{document}